\documentclass[a4paper, 11pt]{article}

\usepackage{jinstpub}

\usepackage{lineno, hyperref}
\usepackage{amsmath}
\usepackage{siunitx}
\usepackage{booktabs, array, makecell}
\usepackage{multirow}
\usepackage{subcaption}
\usepackage{color, soul}
\usepackage{makecell}

\title{\boldmath Spatial resolution improvements with finer-pitch GEMs}

\author[a,b,1]{K.J. Fl\"{o}thner,\note{Corresponding authors.}}
\author[a,1]{L. Scharenberg,}

\author[c,a]{A. Brask}
\author[a]{F. Brunbauer,}
\author[d]{F. Garcia,}
\author[a]{D. Janssens,}
\author[b]{B. Ketzer,}
\author[a,e]{M. Lisowska,}
\author[a,f]{H. Muller,}
\author[a]{E. Oliveri,}
\author[a,g]{G. Orlandini,}
\author[h,a]{D. Pfeiffer,}
\author[a]{L. Ropelewski,}
\author[h,a]{J. Samarati,}
\author[a]{M. van Stenis,}
\author[a]{and R. Veenhof}

\affiliation[a]{European Organization for Nuclear Research (CERN), 1211 Geneva 23, Switzerland}
\affiliation[b]{Helmholtz-Institut f\"{u}r Strahlen- und Kernphysik, University of Bonn, Nu\ss{}allee 14-16, 53115 Bonn, Germany}
\affiliation[c]{Aarhus University, Nordre Ringgade 1, 8000 Aarhus, Denmark}
\affiliation[d]{Helsinki Institute of Physics, P.O. Box 64, FI-00014 University of Helsinki, Finland}
\affiliation[e]{Universit\'{e} Paris-Saclay, F-91191 Gif-sur-Yvette, France}
\affiliation[f]{Physikalisches Institut, University of Bonn, Nu{\ss}allee 12, 53115 Bonn, Germany}
\affiliation[g]{Friedrich-Alexander-Universit{\"a}t Erlangen-N{\"u}rnberg, Schlo\ss{}platz 4, 91054 Erlangen, Germany}
\affiliation[h]{European Spallation Source ERIC (ESS), Box 176, SE-221 00 Lund, Sweden}

\emailAdd{karl.jonathan.floethner@cern.ch, lucian.scharenberg@cern.ch}

\abstract{
	Gas Electron Multipliers (GEMs) are used in many particle physics experiments, employing their `standard' configuration with amplification holes of $\SI{140}{\micro m}$ pitch in a hexagonal pattern.
    However, the collection of the charge cloud from the primary ionisation electrons from the drift region of the detector into the GEM holes affects the position information from the initial interacting particle.
    In this paper, the results from studies with a triple-GEM detector with an X-Y-strip readout anode are presented.
    It is demonstrated that GEMs with a finer hole pitch of here $\SI{90}{\micro m}$ improve the detector's spatial resolution.
    Within these studies, also the impact of the front-end electronics on the spatial resolution was investigated, which is briefly discussed in the paper.
}


\begin{document}

\maketitle
\flushbottom


\section{Introduction}

Detectors based on Gas Electron Multipliers (GEMs) \cite{gem} offer spatial resolutions of better than $\SI{100}{\micro m}$, when equipped with segmented readout anodes of sufficiently small structure size (e.g.\ X-Y-anode strips with $\SI{400}{\micro m}$ pitch \cite{compass-gem}).
Due to the distribution of the created charge in the detector over multiple readout electrodes, spatial resolutions significantly smaller than the structure size can be reached.
This is achieved by using the signal amplitude and reconstruction algorithms such as the centroid/Centre-Of-Gravity (COG) method.
However, in GEM detectors, the distribution is not only affected by the diffusion processes but also by the collection of ionisation and avalanche electrons through the holes of the GEM foils.
Especially, when most probably, only 15 ionisation electrons are created by the interaction of a Minimum-Ionising Particle (MIP) in a $\SI{3}{mm}$ wide drift gap filled with an argon-based gas mixture\footnote{The most probable energy loss of a high-energetic particle in argon is $\SI{1.2}{keV/cm}$ \cite{grupen}, which --- with an average energy loss per ionisation energy of $\SI{26}{eV}$ \cite{grupen} --- corresponds to 46 ionisation electrons per centimetre.}.

\begin{figure}[t!]
    \centering
    \begin{subfigure}{0.45\columnwidth}
        \centering
        \includegraphics[width = \columnwidth]{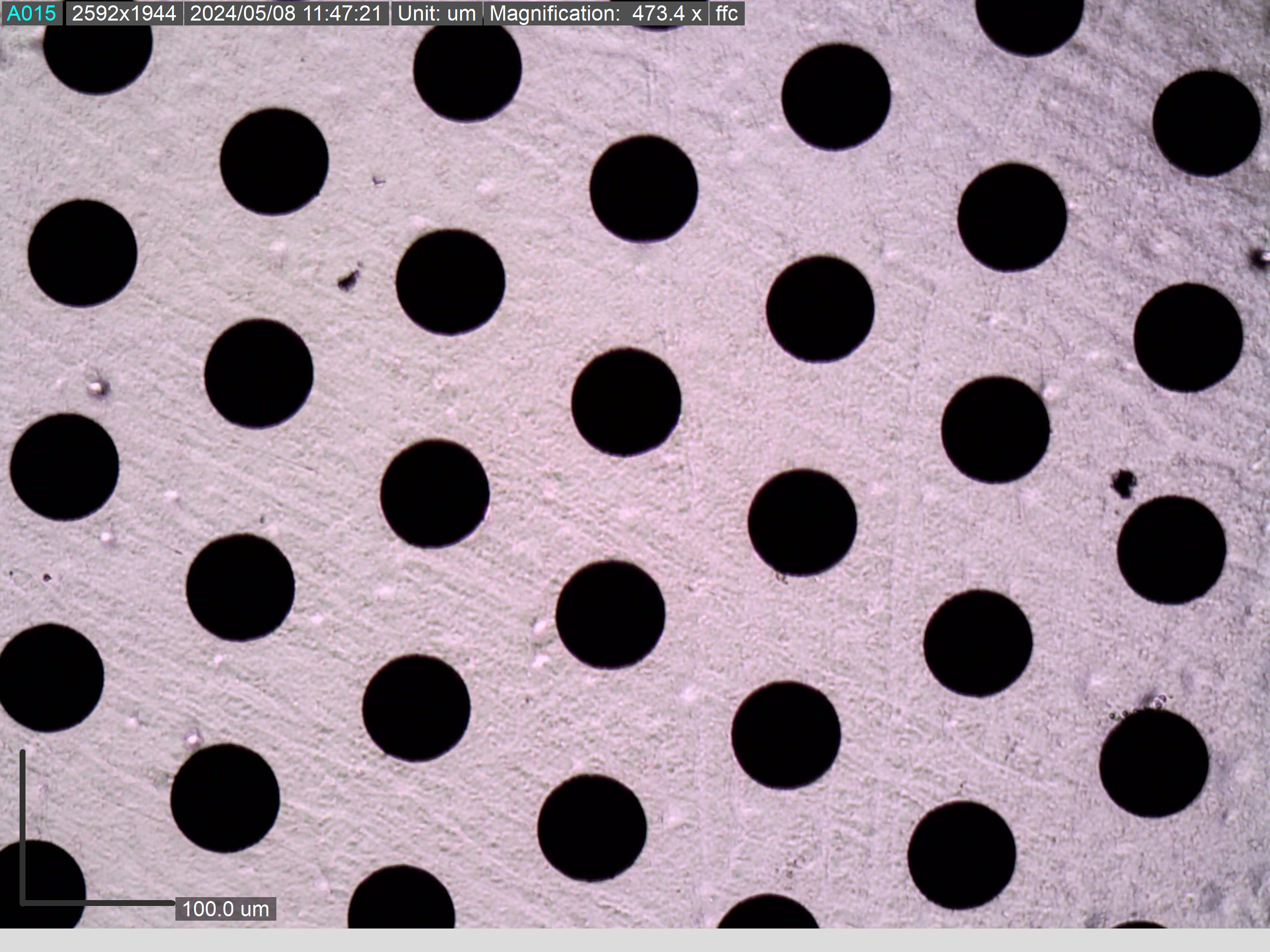}
        \caption{Standard GEM geometry}
        \label{fig:hole-geometries-standard}
    \end{subfigure}
    \hspace{0.05\columnwidth}
    \begin{subfigure}{0.45\columnwidth}
        \centering
        \includegraphics[width = \columnwidth]{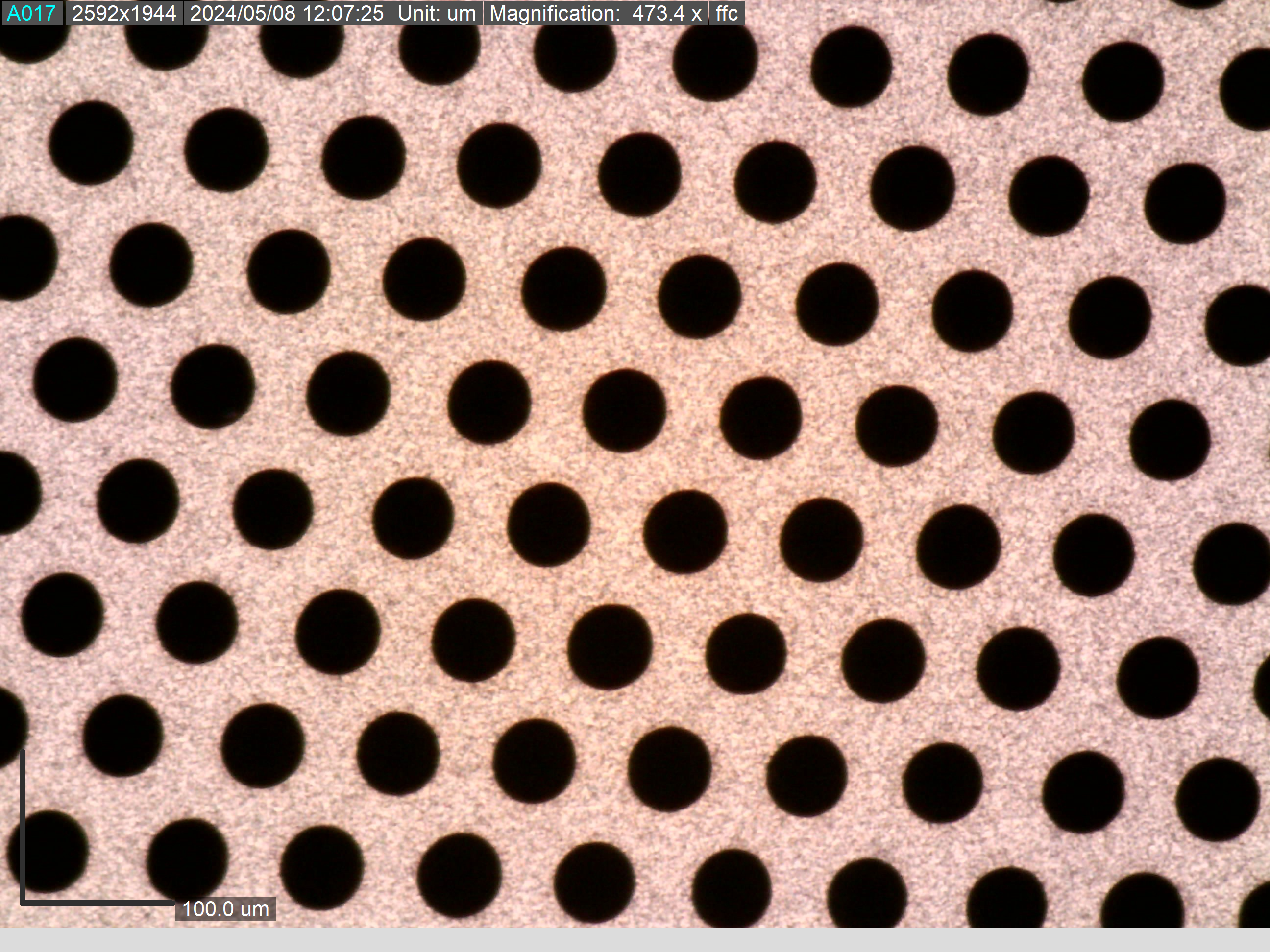}
        \caption{Finer-pitch geometry}
        \label{fig:hole-geometries-fine}
    \end{subfigure}
    \caption{Microscope images of the different GEM foils.}
    \label{fig:hole-geometries}
\end{figure}

In this paper, an approach to improve the spatial resolution of GEM detectors is presented.
Using GEMs with finer hole pitch enables a finer sampling of the primary ionisation electrons during their collection by the holes.
This should improve the preservation of the initial position information and thus the spatial resolution.
The dimensions of these finer-pitch GEMs are shown in Tab.~\ref{tab:gem-dimensions}, in comparison with the standard GEM geometry.
\begin{table}[t]
    \centering
    \caption{Geometrical parameters of the GEM foils.}
    \label{tab:gem-dimensions}
    \begin{tabular}{ccccc}
    \toprule
       & \makecell{Polyimide \\ thickness (\si{\micro m})} & \makecell{Hole \\ pitch (\si{\micro m})} & \makecell{Outer hole \\ diameter (\si{\micro m})} & \makecell{Inner hole \\ diameter (\si{\micro m})}\\
    \midrule
    Standard & 50 & 140 & 70 & 50 \\
    Finer-pitch & 50 & 90 & 55 & 40 \\
    \bottomrule
    \end{tabular}
\end{table}
Microscopic images of the two GEM foil types can be seen in Fig.~\ref{fig:hole-geometries}.

\section{Experimental methods}
\label{sec:experimental-methods}

All measurements have been performed with a COMPASS-like triple-GEM detector \cite{compass-gem} of $\num{10}\times\SI{10}{cm^2}$ active area with an X-Y-readout anode with 256 strips of $\SI{400}{\micro m}$ pitch in each plane.
The charge sharing ratio between the top strips and the bottom strips of this Detector Under Test (DUT) is around 60/40\,\%.
The drift region is $\SI{3}{mm}$ wide, the transfer and induction regions are $\SI{2}{mm}$ wide.
The detector was filled with a gas mixture of Ar/CO\textsubscript{2} (70/30\,\%).
To apply the high voltage to the GEMs, a CAEN HiVolta (DT1415ET) floating channel power supply was used, which allowed the powering of each GEM electrode individually in a stacked divider configuration.

Throughout the measurements, the configuration of the three GEMs within the DUT has been changed, as listed in Tab.~\ref{tab:gem-configurations}.
\begin{table}[t]
    \centering
    \caption{Triple-GEM configurations investigated in the here presented studies, with GEM-1 being exposed to the drift region and GEM-3 being exposed to the induction region and the readout anode.}
    \label{tab:gem-configurations}
    \begin{tabular}{ccccc}
    \toprule
    Configuration & GEM-1 & GEM-2 & GEM-3 & Comment\\
    \midrule
    FP & Finer-pitch & Finer-pitch & Finer-pitch & Full fine-pitch configuration\\
    SG & Standard & Standard & Standard & Standard configuration \\
    Mixed (M.) & Finer-pitch & Standard & Standard & Impact of the first GEM \\
    \bottomrule
    \end{tabular}
\end{table}
In addition to the configurations where the full GEM stack is replaced with finer pitch GEMs, also a configuration with only the first GEM having a finer pitch was investigated, as an increase in the sampling rate should have the largest effect in the charge collection from the drift region.

To understand the impact of the higher granularity on the position reconstruction and the spatial resolution, this DUT was operated during various RD51 test beam campaigns at the H4 beam line of CERN's Super Proton Synchrotron within the RD51 VMM3a/SRS beam telescope \cite{article_vmm-mpgd}.
It consists of three COMPASS-like triple-GEM detectors using the standard geometry with a 50/50\,\% charge-sharing ratio between the anode strip layers.
This allows to reconstruct\footnote{First, the clusters are reconstructed in each strip plane using \emph{vmm-sdat} \cite{vmm-sdat}. This is then followed by the track reconstruction, using \emph{anamicom} \cite{thesis_jona} (\url{https://gitlab.physik.uni- muenchen.de/Jonathan.Bortfeldt/anamicom}), which utilises a Kalman filter \cite{kalman} that is described in detail in \cite{thesis_felix}.} the trajectories of the beam particles --- here $\SI{120}{GeV/\textit{c}}$ muons --- and thus provides the reference position in the DUT.
This reference position $x_\mathrm{ref}$ is then compared with the reconstructed cluster position $x'$ via $\Delta x = x_\mathrm{ref} - x'$, leading to a residual distribution.
The width
\begin{align}
    \sigma_{\Delta x}^2 = w\sigma_\mathrm{core}^2 + (1 - w)\sigma_\mathrm{tail}^2
\end{align}
of this distribution is the basis of the spatial resolution calculation.
It is determined by fitting a double Gaussian function with a core distribution and a tail distribution that are connected by the weighting factor $w$.
Afterwards, the uncertainty from reconstructing the trajectories is subtracted quadratically, as described in \cite{thesis_horvat,thesis_jona}.
The trajectory reconstruction itself is performed on the level of the readout layers of each detector, i.e.\ in the present case with eight layers from the three reference detectors and the DUT.
It is required that at least six of the eight layers participate in the reconstruction process, with the fit of the trajectory being performed only through the reference detectors.
In addition to the spatial resolution, also the detector efficiency can be determined through the track reconstruction.
It is defined as a `hit efficiency' \cite{thesis_jona}
\begin{align}
    \epsilon = \frac{N_\mathrm{hit}}{N_\mathrm{tracks}} = \frac{N_\mathrm{hit}}{N_\mathrm{hit} + N_\mathrm{without}} \ ,
    \label{eq:efficiency}
\end{align}
with $N_\mathrm{hit}$ being the number of tracks with a corresponding cluster in the DUT and $N_\mathrm{without}$ being the number of tracks without a recorded interaction in the DUT.

All the detectors have been read out with the multi-channel VMM3a/SRS front-end electronics \cite{vmm, srs, michael, doro}.
It provides the peak amplitude of the induced charge on each anode strip, which allows to determine the position within the detector using the centroid/Centre-Of-Gravity (COG) method.
The peaking time is adjustable with four discrete settings from $\SI{25}{ns}$ to $\SI{200}{ns}$.
In the presented measurements, $\SI{200}{ns}$ was used, enabling electronics time resolutions of around $\SI{2}{ns}$ \cite{article_vmm-twepp}.
The electronics gain, which can be adjusted with eight discrete settings between $\SI{0.5}{mV/fC}$ and $\SI{16.0}{mV/fC}$, was set to $\SI{6}{mV/fC}$.
In addition, also data sets with $\SI{12}{mV/fC}$ have been taken, illustrating the effects of saturated front-end channels on the results (elaborated in section~\ref{sec:results-electronics}).
The threshold of the front-end electronics was set to values between $\SI{1.7}{fC}$ and $\SI{2.0}{fC}$ for each front-end channel.

\section{Results}
\label{sec:results}

In the following, the results obtained for the three different detector configurations (Tab.~\ref{tab:gem-configurations}) are shown.
These configurations have been used with different voltage settings.
At first, the so-called `COMPASS settings' \cite{compass-gem} were used: the voltage ratios between the drift region, GEM-1, the first transfer gap, GEM-2, the second transfer gap, GEM-3 and the induction gap are $1/0.55/1/0.5/1/0.45/1$.
This was only used for the GEM stack configurations FP and SG.
In addition, for all three stack configurations, also the voltage across GEM-1 was varied in the range from $\SI{300}{V}$ to $\SI{420}{V}$.
The other voltages were kept at $\SI{740}{V}$ across the drift, transfer and induction fields, with $\SI{370}{V}$ across GEM-2 and $\SI{333}{V}$ across GEM-3, corresponding to the nominal COMPASS settings \cite{compass-gem}.

The change in the operating voltage corresponds to a change in the effective detector gain.
It should be noted that the data points are not plotted against gain values, but against the most probable value from the energy-loss (Landau) distribution, given as the total measured cluster charge in ADC values.
To cover the entire detector gain range, the peak positions in the region of full detector efficiency were used to extrapolate down towards the regions with lower efficiency.
From measuring the current on the anode strips of the DUT in the laboratory, using a \textsuperscript{55}Fe radioactive source, it could be deduced that an ADC value of around $\num{1000}$ --- at $\SI{6}{mV/fC}$ electronics gain --- corresponds to an effective detector gain of around $\num{4e4}$.

\subsection{Cluster size}

\begin{figure}[t!]
    \centering
    \begin{subfigure}{0.45\columnwidth}
        \centering
        \includegraphics[width = 0.884956\columnwidth]{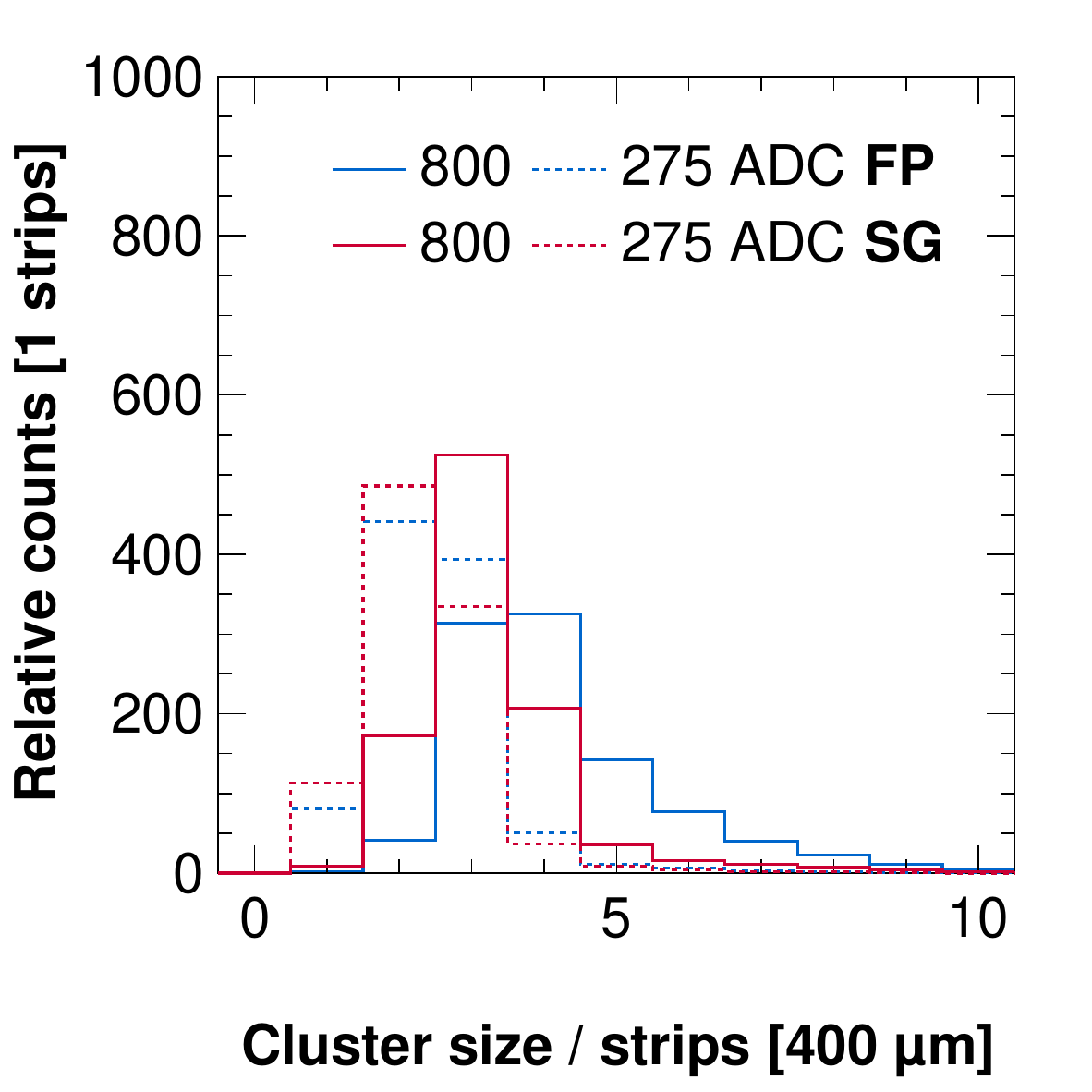}
        \caption{Cluster size distributions}
        \label{fig:results-cluster-size-compass-distribution}
    \end{subfigure}
    \hspace{0.05\columnwidth}
    \begin{subfigure}{0.45\columnwidth}
        \centering
        \includegraphics[width = 0.884956\columnwidth]{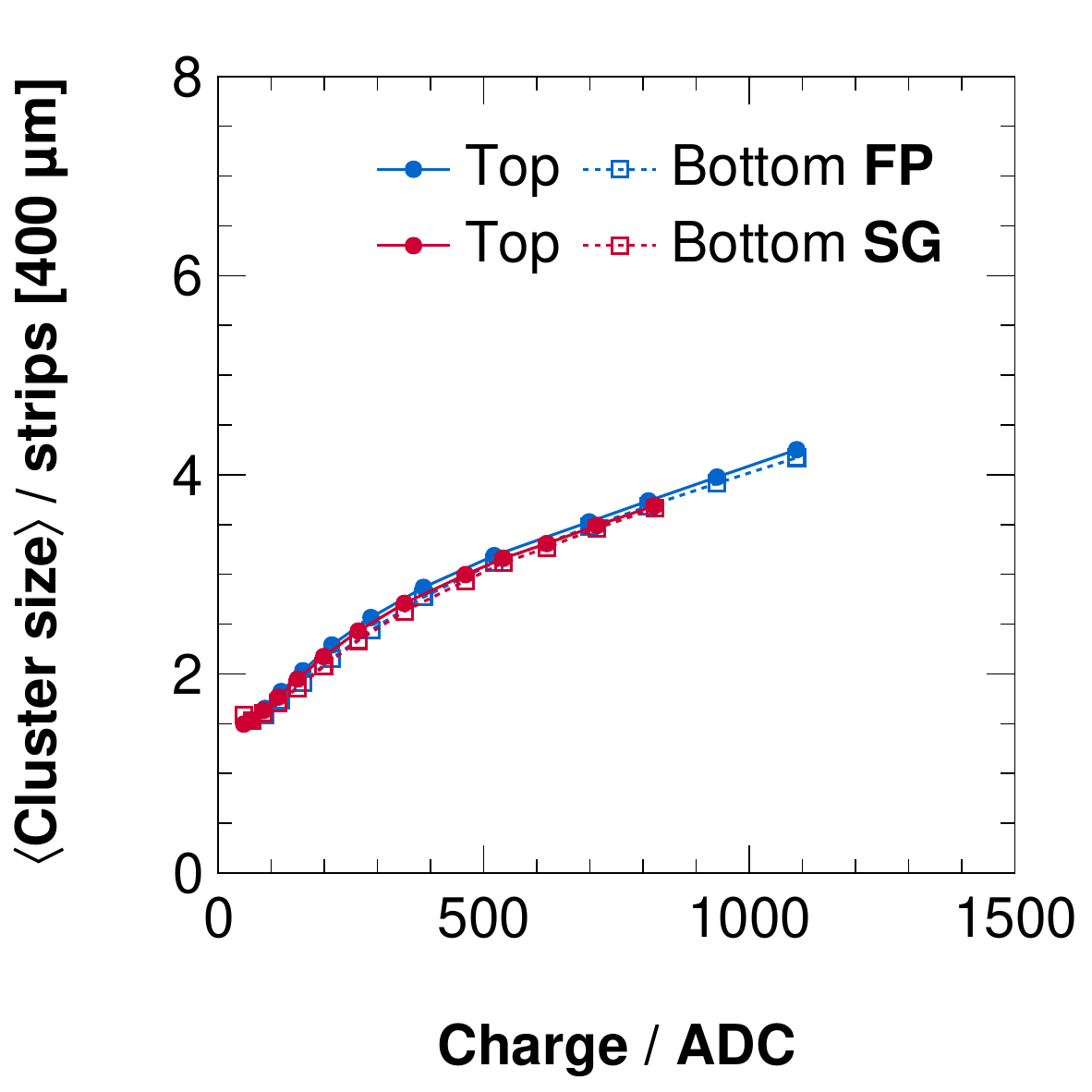}
        \caption{Average cluster size}
        \label{fig:results-cluster-size-compass-trend}
    \end{subfigure}
    \caption{In (a), selected cluster size distributions are shown for the top strips at two different detector gains.
        In (b), the average of these distributions is shown for various gain values, both for the top and bottom strips, as well as for the FP and the SG configuration, using the COMPASS settings.}
    \label{fig:results-cluster-size-compass}
\end{figure}

\begin{figure}[t!]
    \centering
    \begin{subfigure}{0.45\columnwidth}
        \centering
        \includegraphics[width = 0.884956\columnwidth]{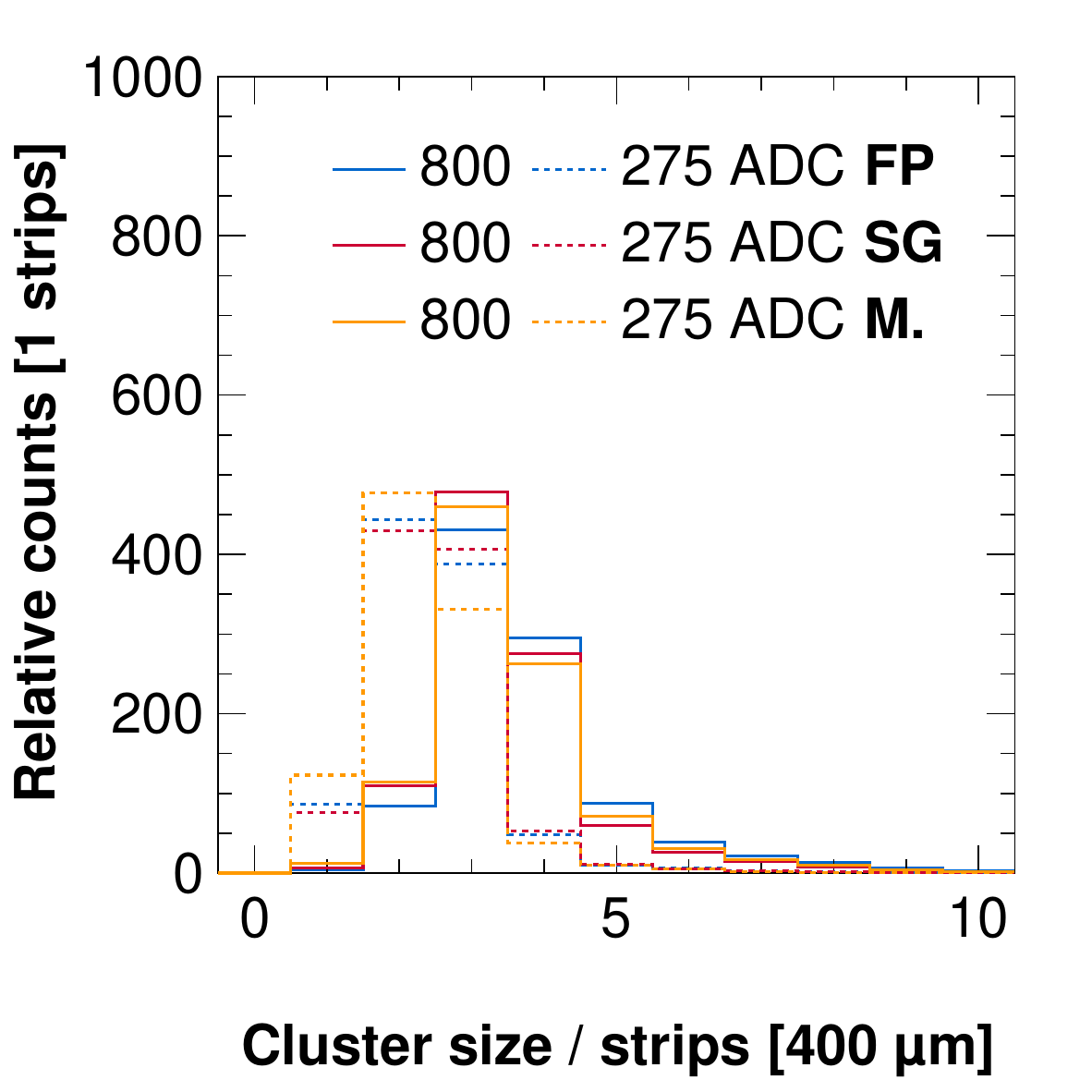}
        \caption{Cluster size distributions}
        \label{fig:results-cluster-size-g1-distribution}
    \end{subfigure}
    \hspace{0.05\columnwidth}
    \begin{subfigure}{0.45\columnwidth}
        \centering
        \includegraphics[width = 0.884956\columnwidth]{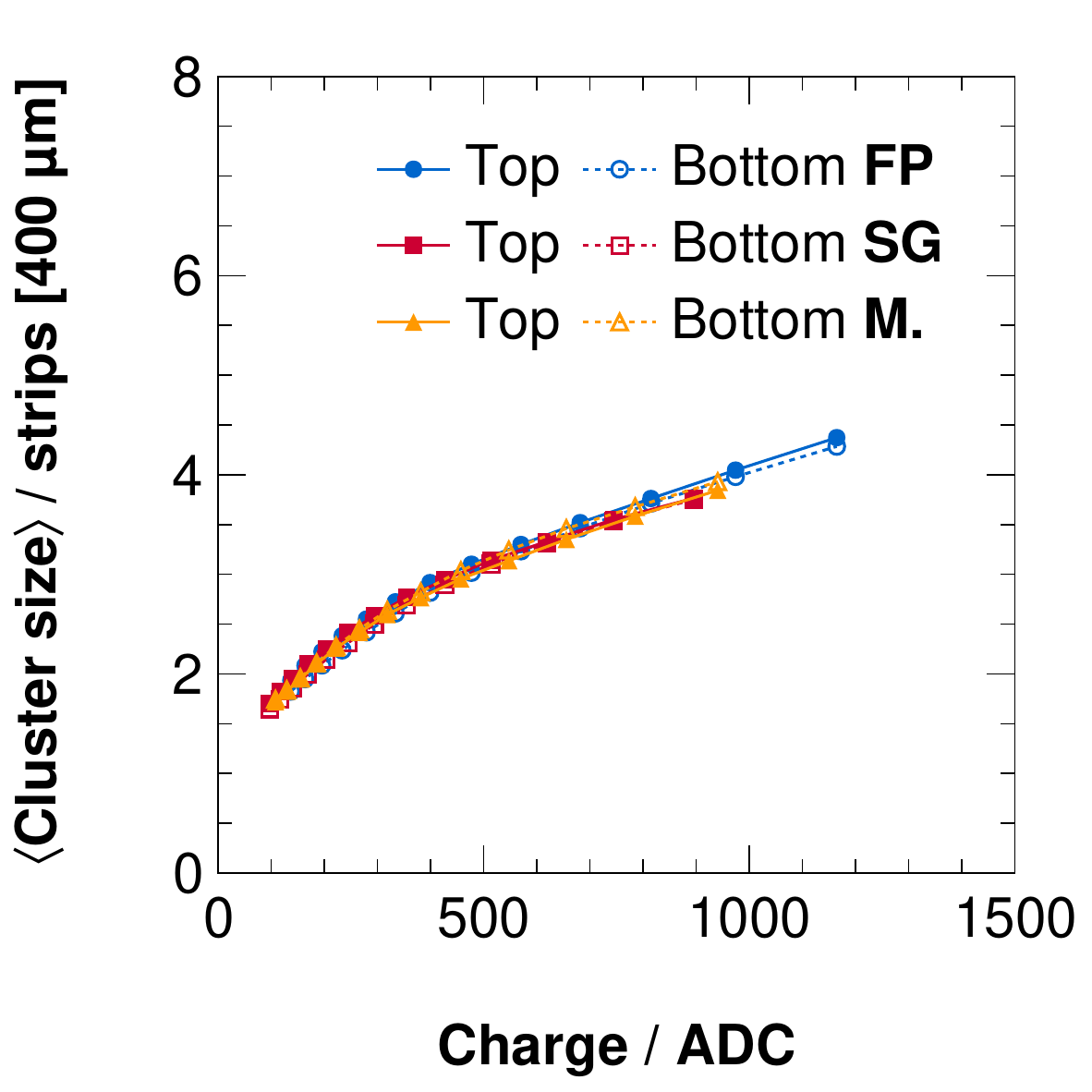}
        \caption{Average cluster size}
        \label{fig:results-cluster-size-g1-trend}
    \end{subfigure}
    \caption{In (a), selected cluster size distributions are shown for the top strips at two different detector gains.
        In (b), the average of these distributions is shown for various gain values, both for the top and bottom strips for the GEM-1 voltage scan.}
    \label{fig:results-cluster-size-g1}
\end{figure}

Before investigating the spatial resolution behaviour, the cluster size, i.e.\ the number of channels above THL within a cluster, is presented.
For the COMPASS settings, the results are shown in Fig.~\ref{fig:results-cluster-size-compass}.
For varying the voltage across GEM-1, the results are shown in Fig.~\ref{fig:results-cluster-size-g1}.
In both figures, examples of the cluster size distribution are shown for different most probable charge values, i.e.\ different detector gains.
In the trend of the average cluster size (Figs.~\ref{fig:results-cluster-size-compass-trend} and \ref{fig:results-cluster-size-g1-trend}), a slight `kink' can be observed, located at around $\SI{350}{ADCs}$.
When comparing this point with the efficiency behaviour (e.g.\ Figs.~\ref{fig:results-gain} and \ref{fig:results-gem1}) this is the same value at which the detectors become fully efficient.

\subsection{Spatial resolution}

\begin{figure}[t!]
    \centering
    \begin{subfigure}{0.45\columnwidth}
        \centering
        \includegraphics[width = \columnwidth]{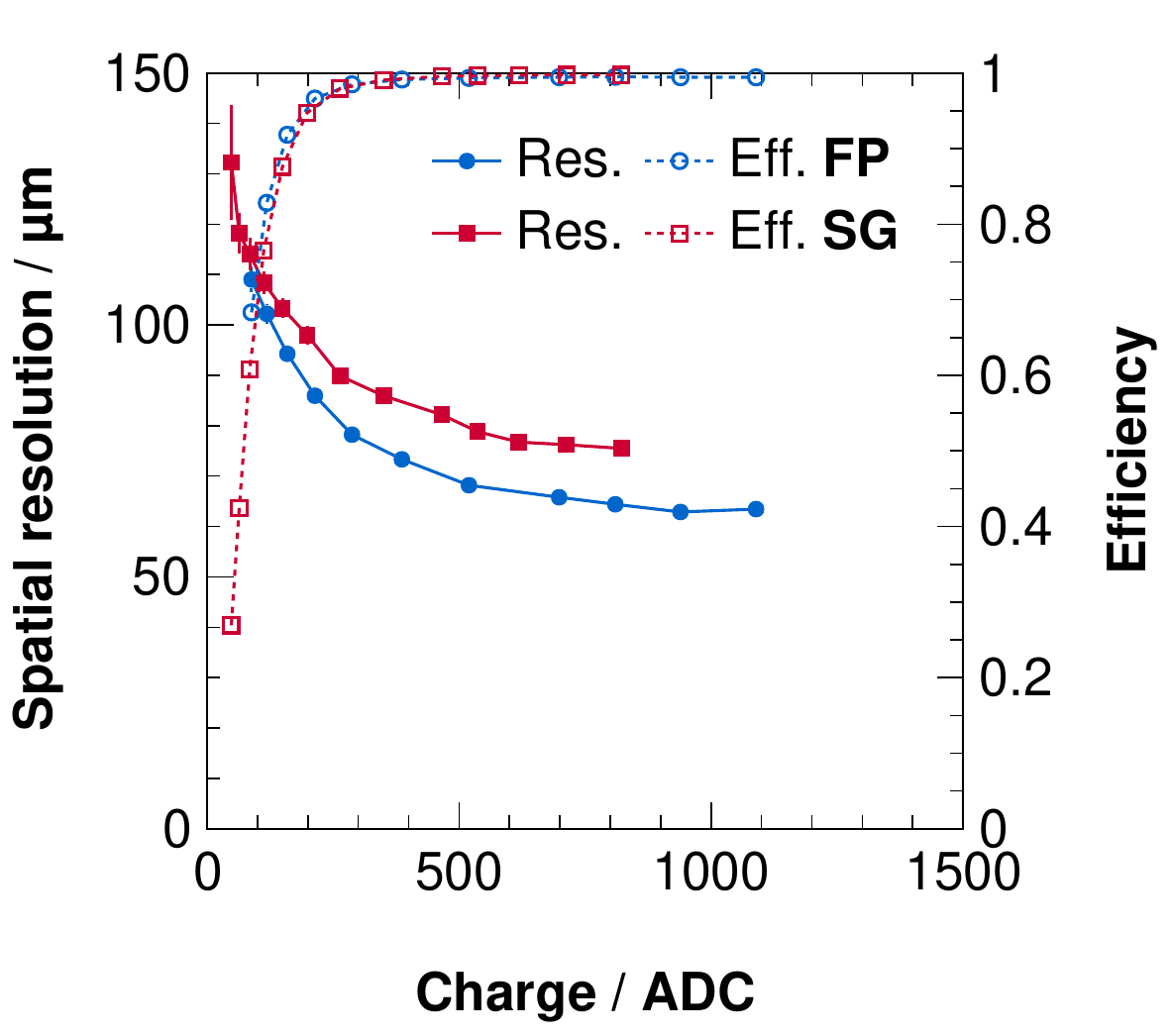}
        \caption{Top strips}
        \label{fig:results-gain-top}
    \end{subfigure}
    \hspace{0.05\columnwidth}
    \begin{subfigure}{0.45\columnwidth}
        \centering
        \includegraphics[width = \columnwidth]{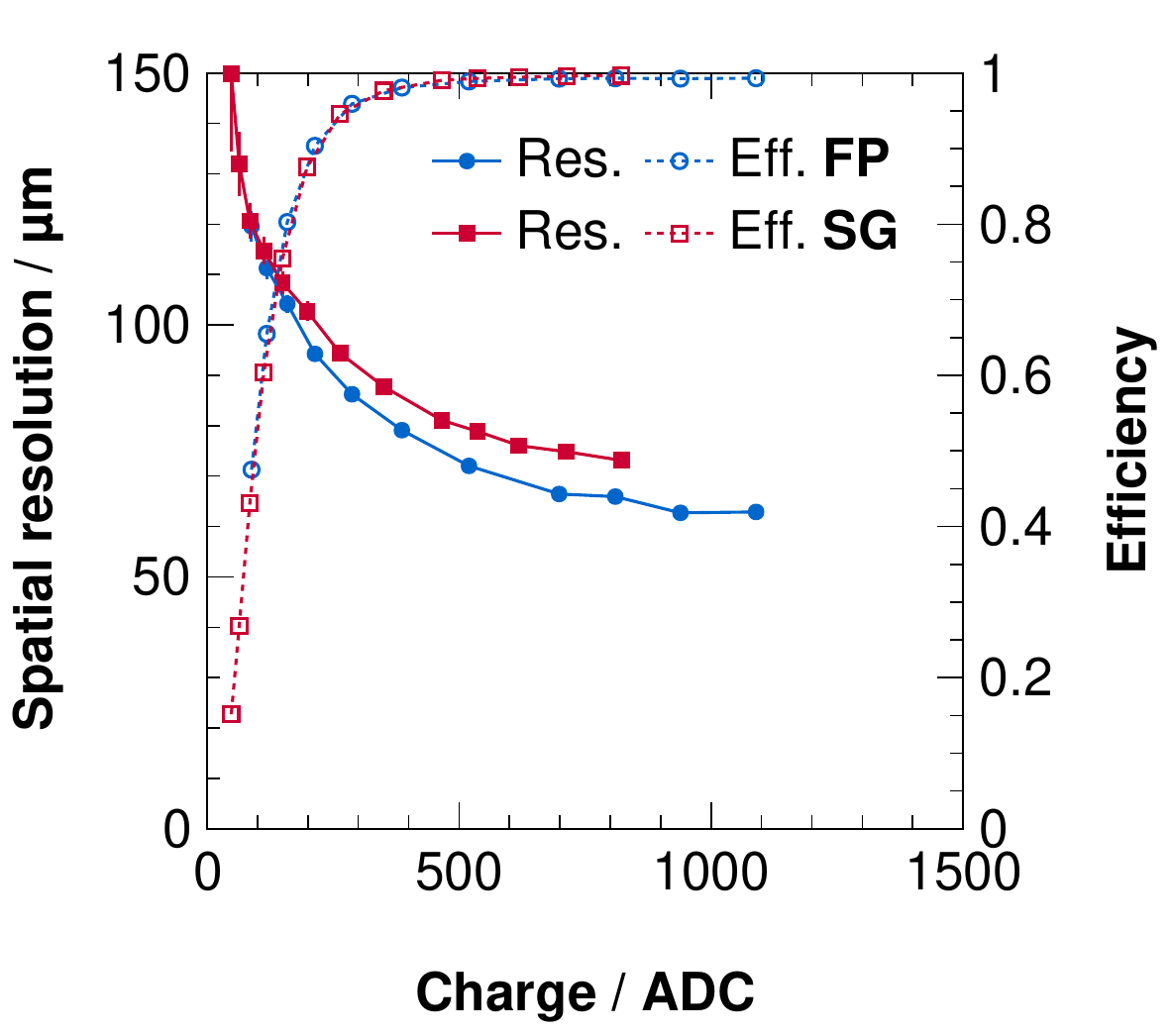}
        \caption{Bottom strips}
        \label{fig:results-gain-bottom}
    \end{subfigure}
    \caption{Spatial resolution (Res.) and detector efficiency (Eff.), comparing the Finer Pitch (FP) GEMs and ones with Standard Geometry (SG), depending on the detector gain.}
    \label{fig:results-gain}
\end{figure}

In the following, the observed spatial resolution is presented for the three different detector configurations, starting with the COMPASS settings.
The results are shown in Fig.~\ref{fig:results-gain}, for each strip layer individually.
It can be seen that by using finer-pitch GEMs, the spatial resolution improves by around $\SI{10}{\micro m}$ or approximately $\SI{15}{\percent}$, especially towards higher gains ($>\SI{200}{\text{ADC values}}$, corresponding to a gain of around $\num{e4}$ per plane).

\begin{figure}[t!]
    \centering
    \begin{subfigure}{0.45\columnwidth}
        \centering
        \includegraphics[width = \columnwidth]{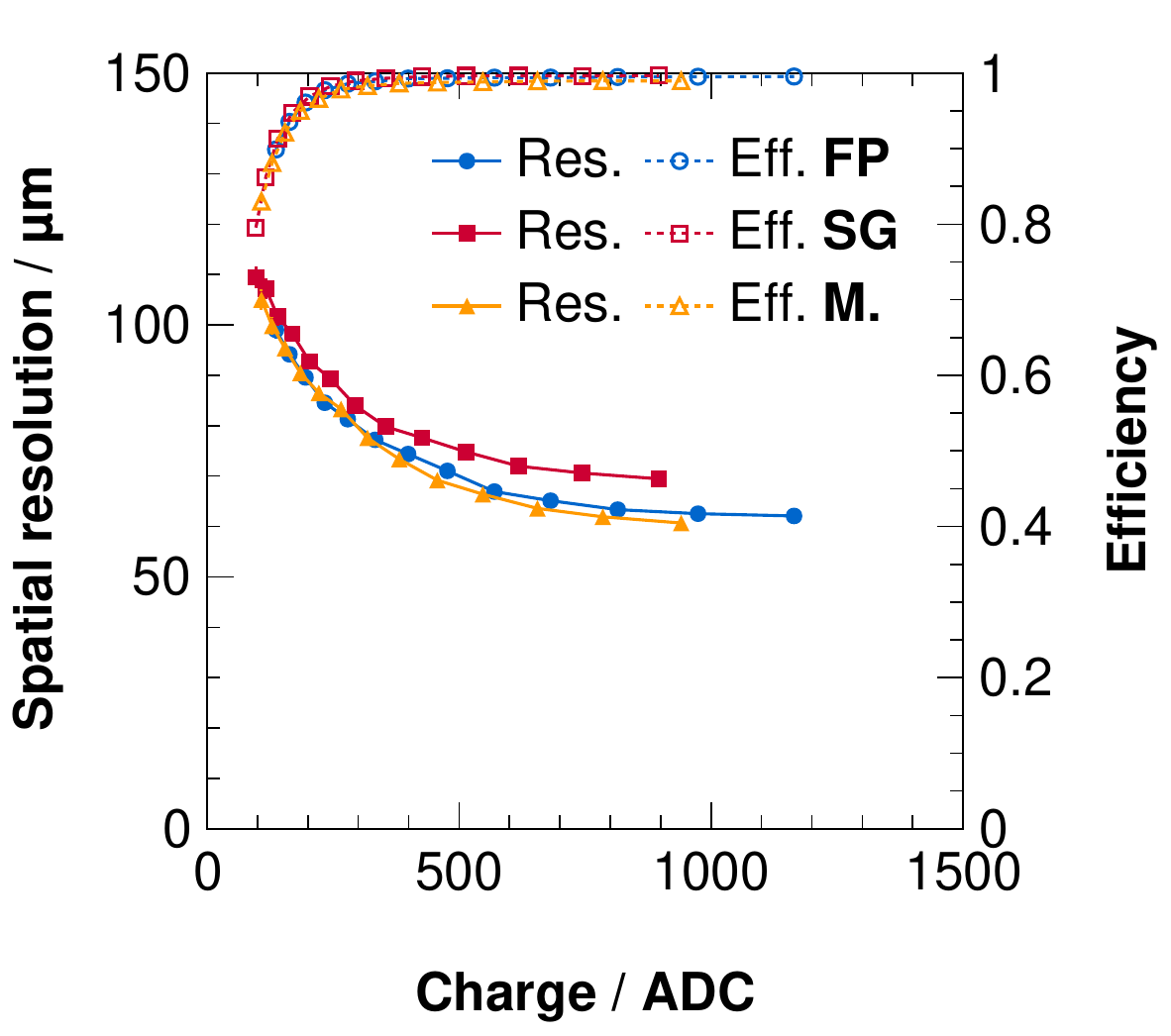}
        \caption{Top strips}
        \label{fig:results-gem1-top}
    \end{subfigure}
    \hspace{0.05\columnwidth}
    \begin{subfigure}{0.45\columnwidth}
        \centering
        \includegraphics[width = \columnwidth]{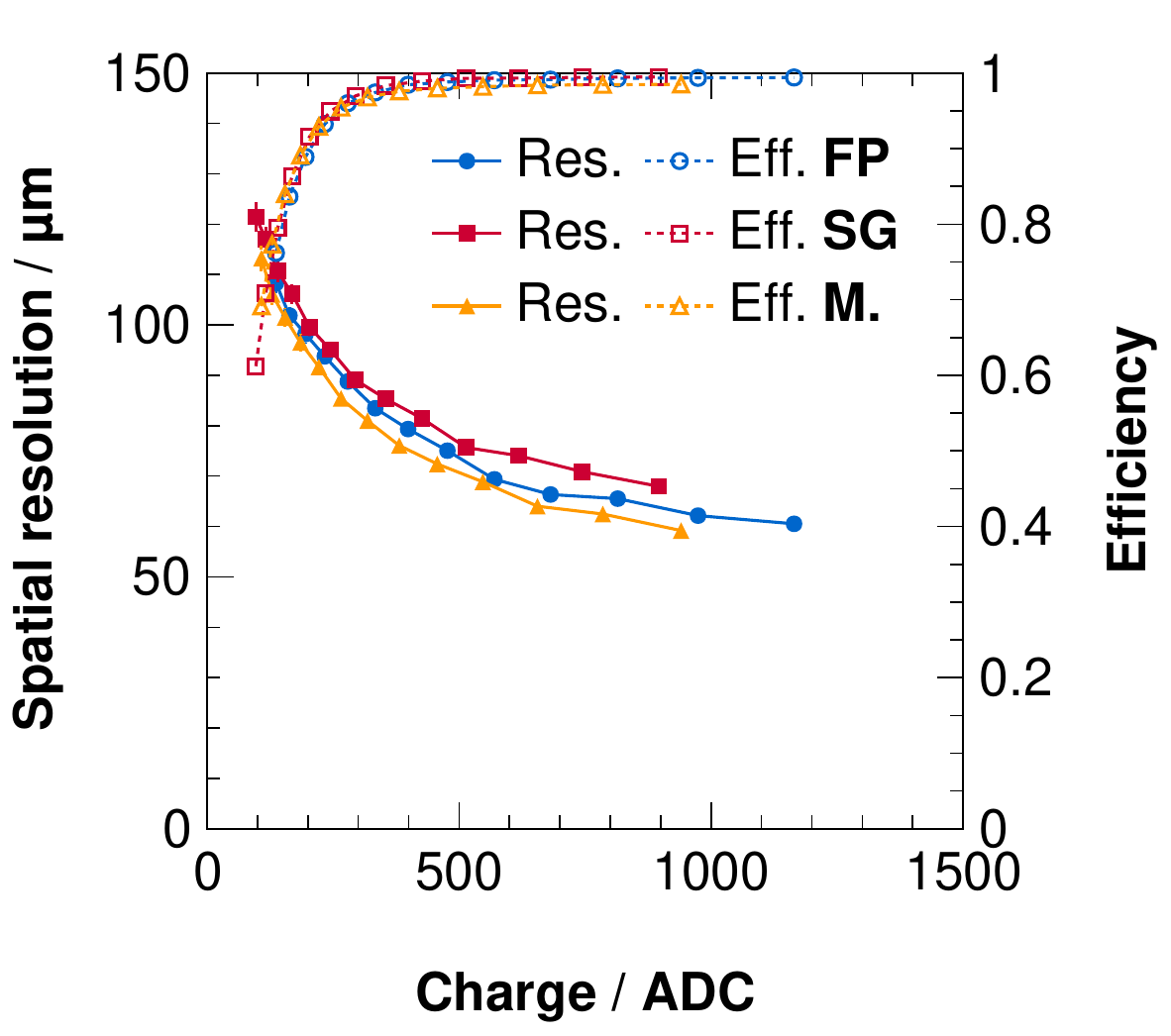}
        \caption{Bottom strips}
        \label{fig:results-gem1-bottom}
    \end{subfigure}
    \caption{Spatial resolution (Res.) and detector efficiency (Eff.), comparing all three configurations, depending on the detector gain achieved by changing the voltage across the first GEM.}
    \label{fig:results-gem1}
\end{figure}

The same behaviour is observed for all three stack configurations (Tab.~\ref{tab:gem-configurations}) when the voltage across GEM-1 is varied (Fig.~\ref{fig:results-gem1}).
However, the improvement between FP and SG configuration is less pronounced than for the COMPASS settings.
It can be also seen that the mixed configuration, using a finer-pitch GEM only as a first foil in the detector, provides the best results.
A possible explanation for this might be found in the way the gain was varied.
By changing only the voltage across GEM-1, while keeping the potential difference between the top electrode of GEM-1 and the cathode, as well as the bottom electrode of GEM-1 and the top electrode of GEM-2 at the same values, the voltage ratio changes, compared to the COMPASS settings.
Hence, despite achieving the same gain in terms of measured charge, the focusing effects from the field lines, i.e.\ the charge collection and extraction differ depending on the GEM parameters \cite{ottnad}.
While the fields have been optimised for the standard geometry GEMs, they might not be optimal for the finer-pitch GEMs thus affecting the spatial resolution.

\begin{figure}[t!]
    \centering
    \begin{subfigure}{0.45\columnwidth}
        \centering
        \includegraphics[width = 0.884956\columnwidth]{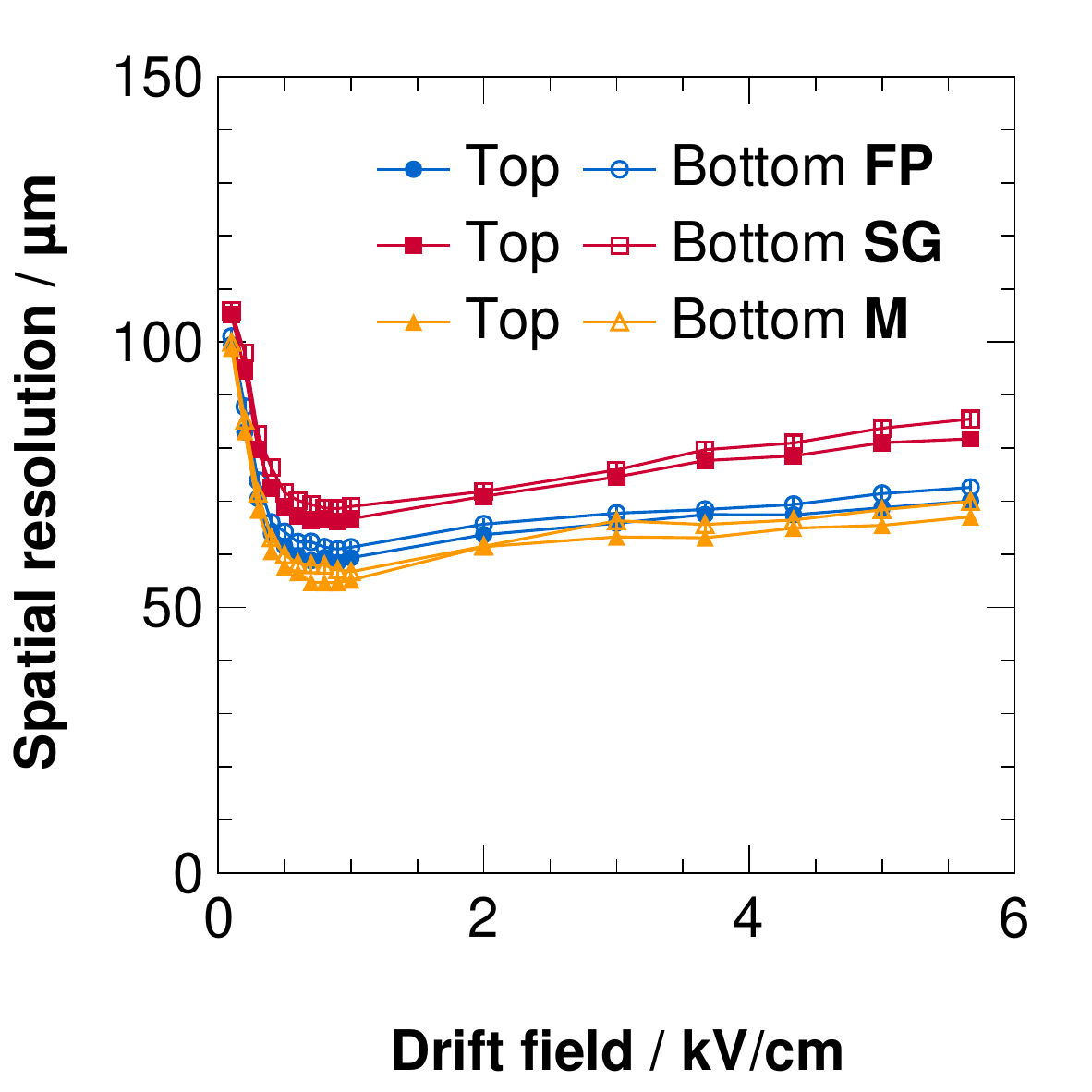}
        \caption{Top strips}
        \label{fig:results-drift-data}
    \end{subfigure}
    \hspace{0.05\columnwidth}
    \begin{subfigure}{0.45\columnwidth}
        \centering
        \includegraphics[width = 0.884956\columnwidth]{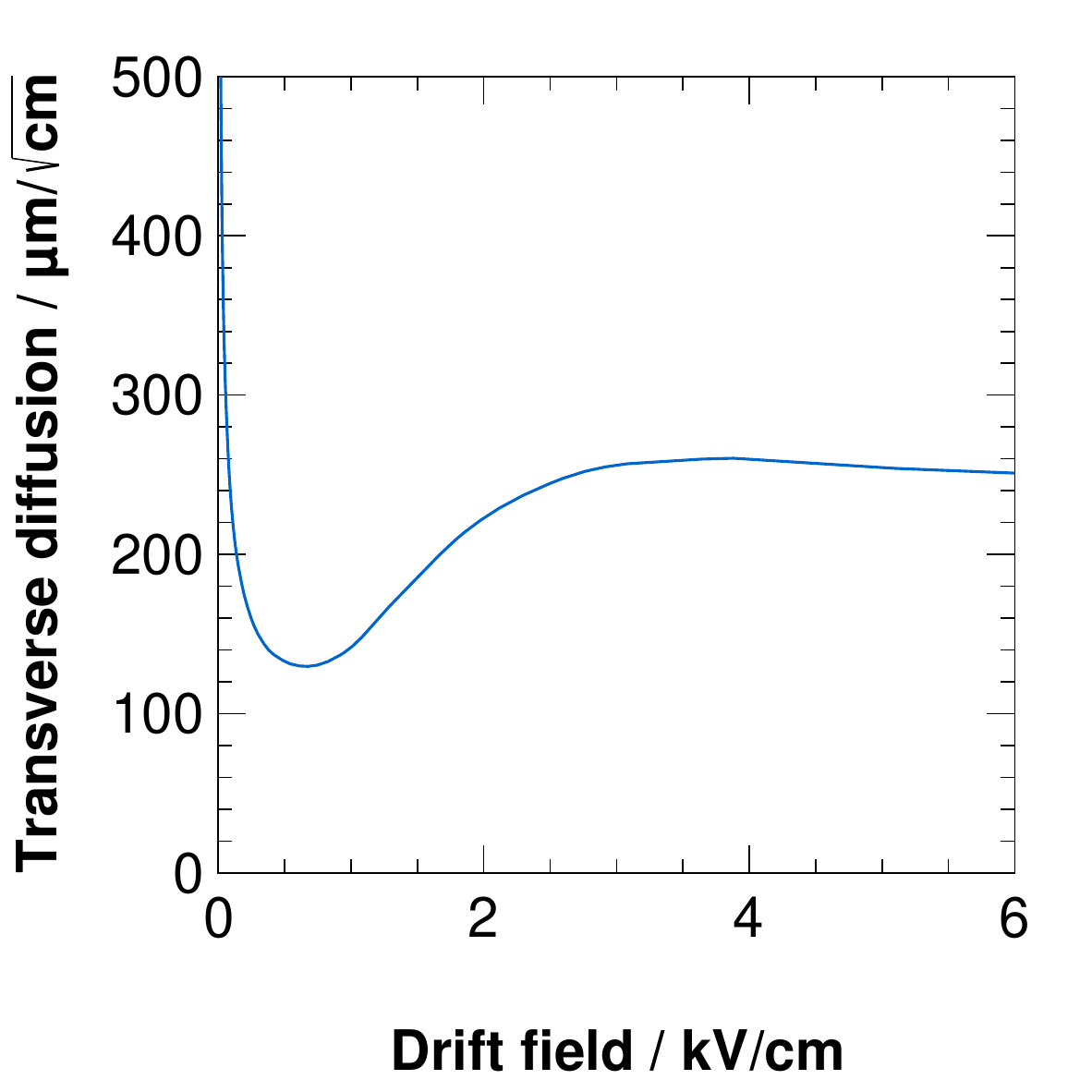}
        \caption{Bottom strips}
        \label{fig:magboltz}
    \end{subfigure}
    \caption{In (a), the spatial resolution (Res.) and detector efficiency (Eff.), depending on the drift field is shown.
        In (b), the transverse diffusion in Ar/CO\textsubscript{2} (70/30\,\%), calculated with Magboltz \cite{magboltz}, depending on the drift field is shown.}
    \label{fig:results-drift}
\end{figure}

In addition to the gain dependence, also the dependence of the spatial resolution on the drift field was investigated.
The results are shown in Fig.~\ref{fig:results-drift}.
It can be seen that the best spatial resolution is obtained at drift fields of around $\SI{0.7}{kV/cm}$.
This can be related to the transversal diffusion of the electron cloud (Fig.~\ref{fig:magboltz}), indicating that the minimal spatial resolution is correlated with the smallest transverse diffusion.
It can be also seen that the difference in spatial resolution gets larger towards higher drift fields, especially when comparing the FP/mixed with the SG configuration.
This could be explained by the focusing of the field lines, leading to primary electrons ending up on the copper electrodes of the GEM foils.
With the higher hole density of the first GEM in the FP/mixed configuration, this effect is less pronounced.

\subsection{Excursus --- effects of saturated front-end channels}
\label{sec:results-electronics}

\begin{figure}[t!]
    \centering
    \begin{subfigure}{0.45\columnwidth}
        \centering
        \includegraphics[width = \columnwidth]{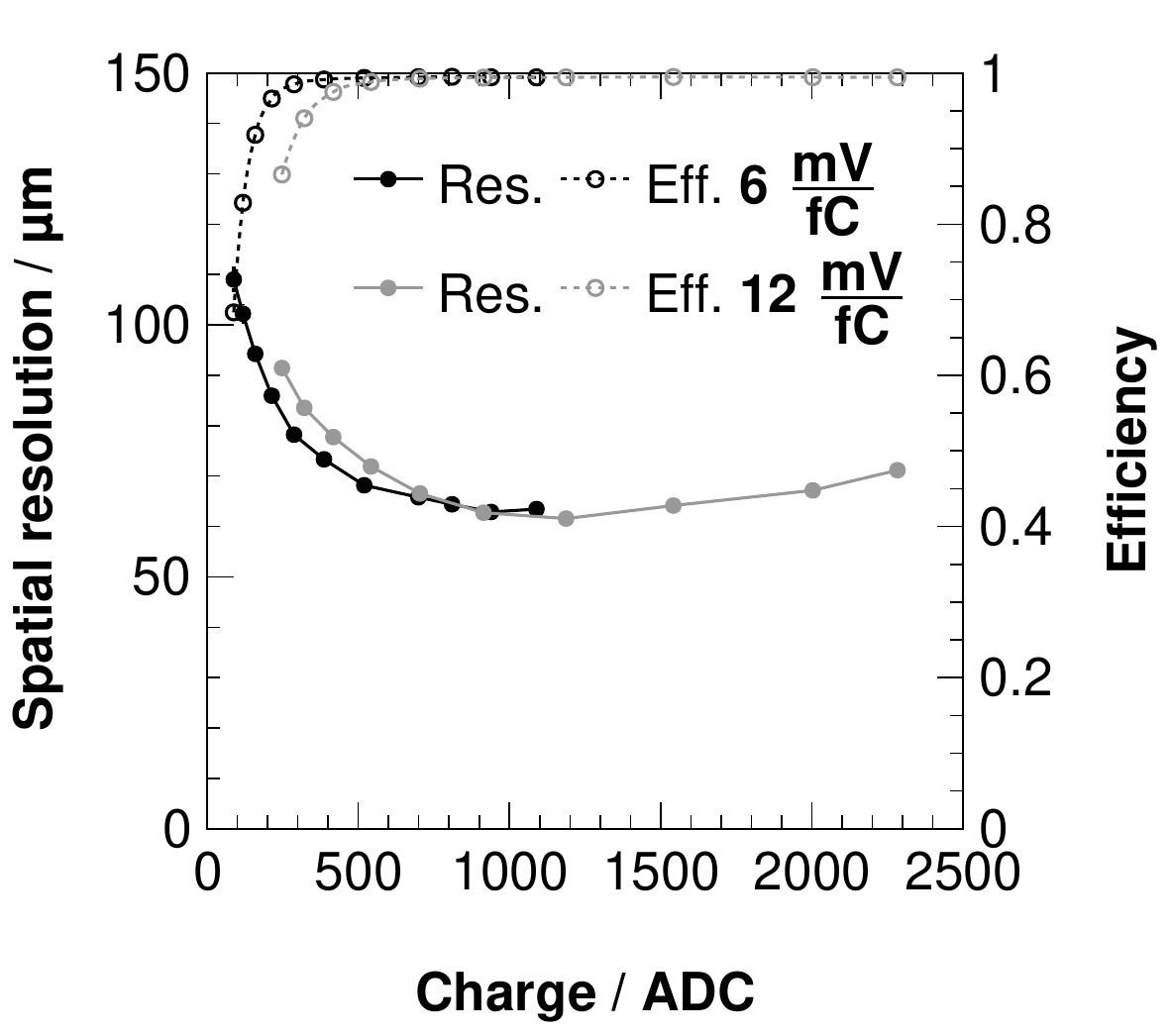}
        \caption{Different electronics gains}
        \label{fig:results-saturation-default}
    \end{subfigure}
    \hspace{0.05\columnwidth}
    \begin{subfigure}{0.45\columnwidth}
        \centering
        \includegraphics[width = \columnwidth]{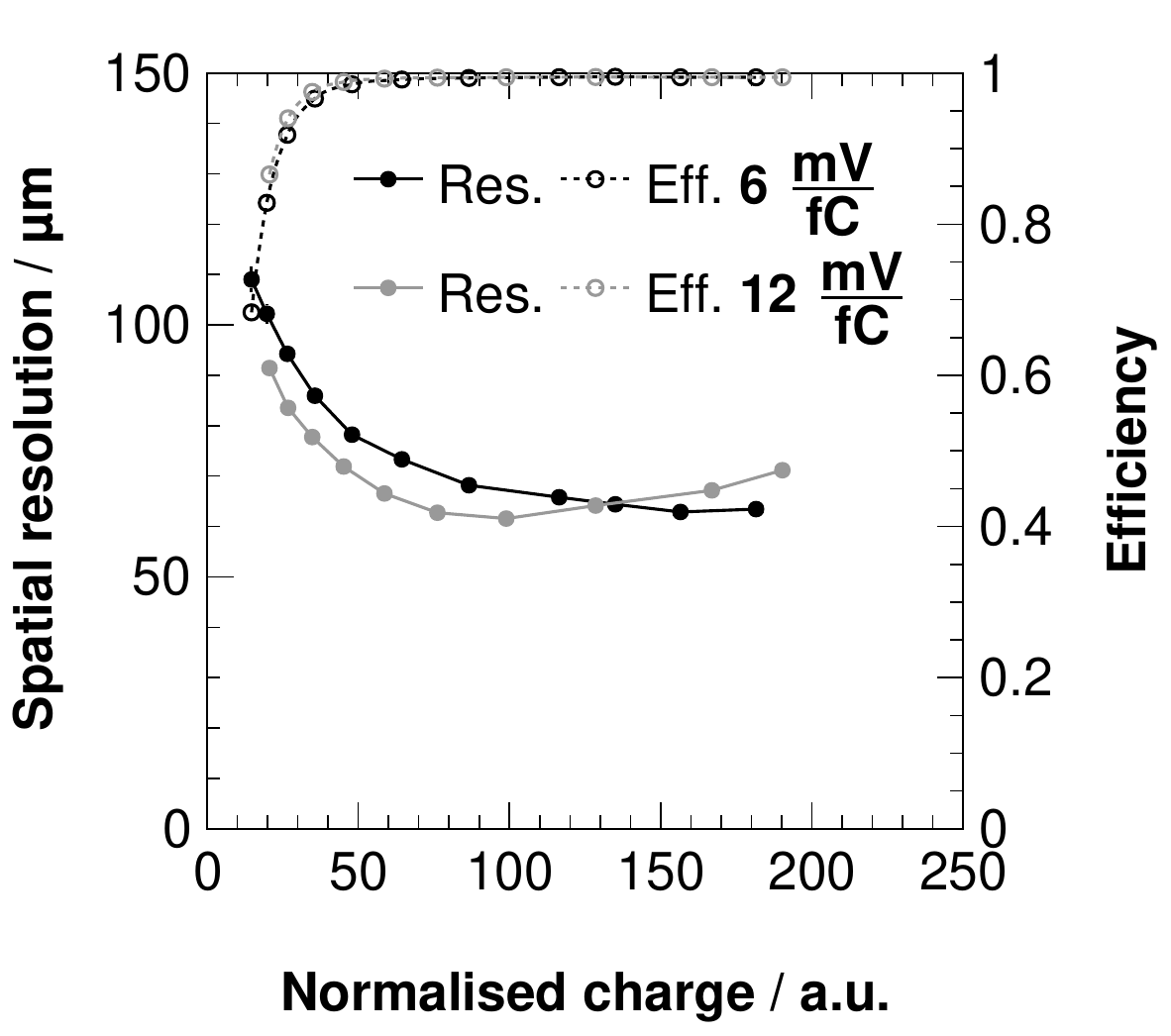}
        \caption{Charge normalised by electronics gain}
        \label{fig:results-saturation-scaled}
    \end{subfigure}
    \caption{Spatial resolution (Res.) and detector efficiency (Eff.), recorded for the COMPASS settings at different electronics gains, showing only the top strips.
        In (a), the trend is shown, purely on the recorded ADC value, while in (b) the same behaviour is shown, with the charge being normalised by the electronics gain.}
    \label{fig:results-saturation}
\end{figure}

During the measurements, it was found how the front-end electronics --- specifically saturated front-end channels --- can impact the results.
Especially, when operating the DUT at high detector gains in combination with a higher electronics gain of $\SI{12}{mV/fC}$, it was observed that the spatial resolution decreases when increasing the detector gain (Fig.~\ref{fig:results-saturation}).
Although this may seem like a fairly obvious statement, the impact on the investigated detector behaviour is not negligible, even in regions of the dynamic range where on the first sight no saturation is observed.

This is illustrated in Fig.~\ref{fig:results-saturation-scaled}.
There, the results are normalised according to their electronics gain.
The efficiency behaviour now matches for both gain cases and reflects only the detector behaviour --- as expected, as it is a pure counting experiment.
The spatial resolution behaviour however is better for the $\SI{12}{mV/fC}$ data at low detector gains, as more of the dynamic range of the ADC is used, i.e.\ the impact of the non-differential linearity of the ADC\footnote{Although the ADC values are sent out as 10-bit values, the effective resolution of the charge ADC is between 6-bit and 7-bit.} is less significant.
On the other hand, it can be seen that towards higher detector gains the spatial resolution decreases again, due to more channels going into saturation.

\begin{figure}[t!]
    \centering
    \begin{subfigure}{0.45\columnwidth}
        \centering
        \includegraphics[width = \columnwidth]{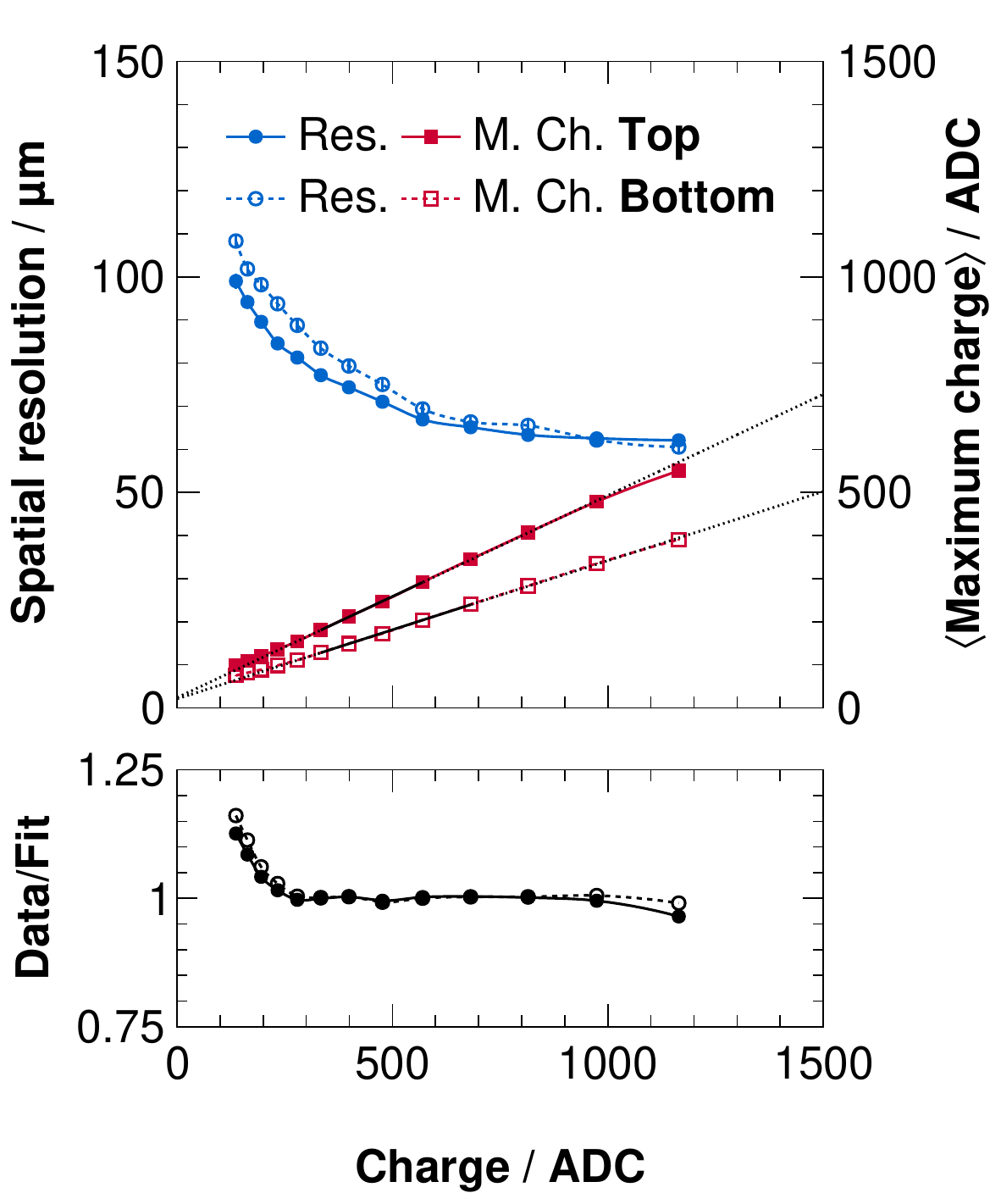}
        \caption{$\SI{6}{mV/fC}$}
        \label{fig:results-saturation-overlay-6mV}
    \end{subfigure}
    \hspace{0.05\columnwidth}
    \begin{subfigure}{0.45\columnwidth}
        \centering
        \includegraphics[width = \columnwidth]{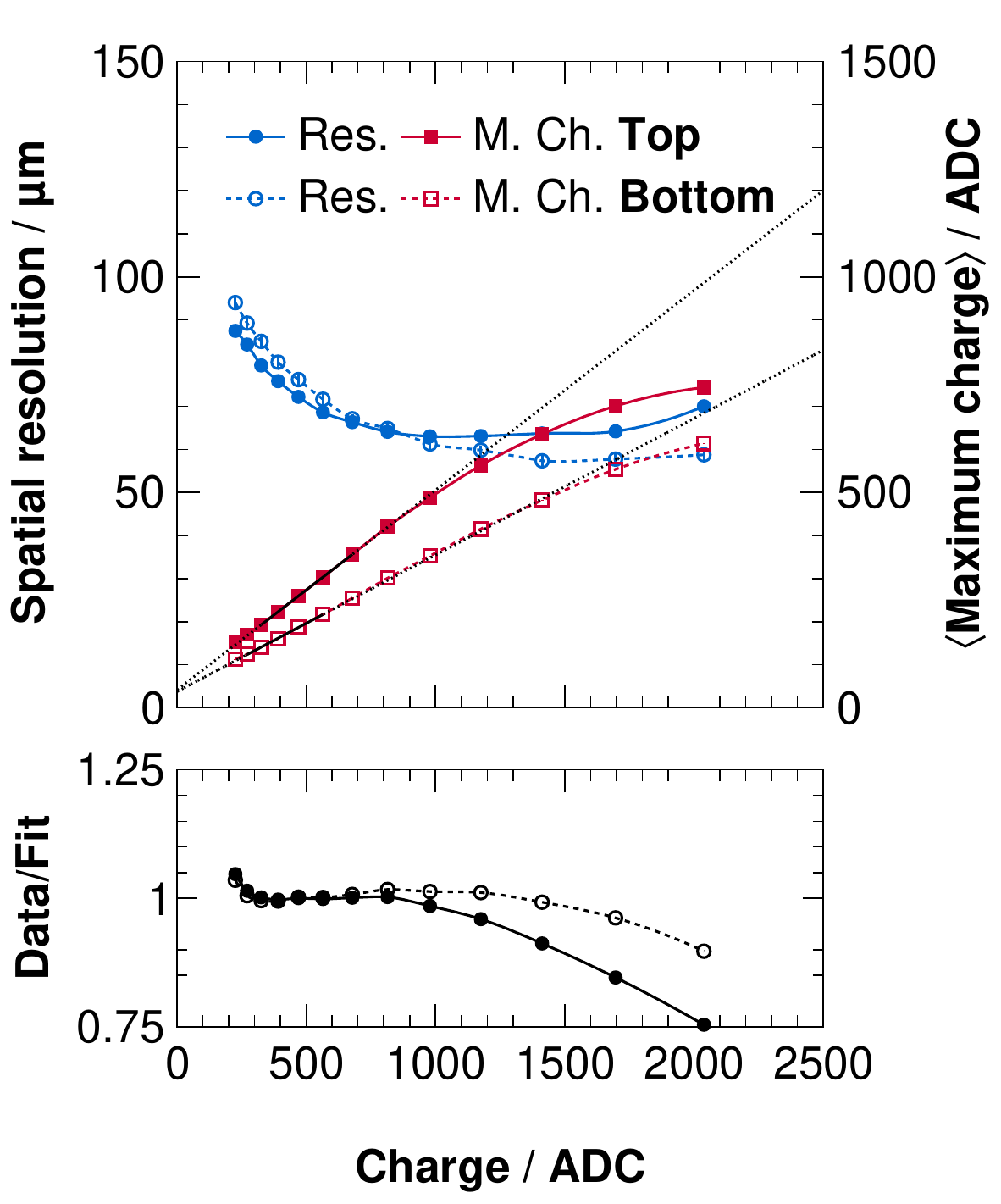}
        \caption{$\SI{12}{mV/fC}$}
        \label{fig:results-saturation-overlay-12mV}
    \end{subfigure}
    \caption{Spatial resolution (Res.) and average maximum charge (M.\ Ch.), recorded for the GEM-1 variation at different electronics gains.
        The fit of the linear function to the maximum charge data points is only performed in the region with a solid fit line.
        The dashed line is an extrapolation of the fit.}
    \label{fig:results-saturation-overlay}
\end{figure}

This is illustrated in Fig.~\ref{fig:results-saturation-overlay}.
The spatial resolution behaviour is shown for both strip planes at different electronics gains, determined for the GEM-1 scan.
In addition, the average maximum charge is shown:
from each reconstructed cluster in the DUT, the channel with the largest amplitude is identified, followed by averaging all these amplitudes.
For this quantity, a flattening towards higher gains can be observed, especially at $\SI{12}{mV/fC}$.
The flatting is also less pronounced for the bottom strips as they collect only $\SI{40}{\percent}$ of the charge, while the top strips collect $\SI{60}{\percent}$, which also explains the numerical difference between the top and bottom strips.
Fitted to the charge curves are linear functions.
To determine the best fit-region --- meaning a region with the most linear behaviour --- the fit range was varied, with the reduced $\chi^2$ being determined for each region and the fit with the reduced $\chi^2$ closest to 1 being selected.
Underneath these graphs, the ratio plot between the measured data for the charge curves and the fit to these data is shown.
This helps to get an indication of when the linearity of the charge behaviour is lost, due to the saturation of front-end channels.

It also shows the transition in the saturation behaviour and the use of the dynamic range.
Even in cases where no saturation is expected at first sight --- specifically the points of the $\SI{6}{mV/fC}$ data set at the highest gain --- it can be seen from the deviation from linearity that the saturation of the front-end channels takes place and slightly decreases the spatial resolution results:
the top strips have a worse spatial resolution than the bottom strips.
But because the approaching of the spatial resolution of the top and bottom strips starts already at lower gains, even here the results are affected by the electronics.
This shows the importance of matching electronics and detector settings for optimal results.
Furthermore, it displays the benefits of readout electronics with adjustable gain to match the detector settings.

\section{Summary and outlook}

In this paper, a novel approach to improve the spatial resolution of triple-GEM detectors with strip readout ($\SI{400}{\micro m}$ pitch) has been described.
Instead of using Gas Electron Multipliers with the standard hole geometry with $\SI{140}{\micro m}$ pitch holes, GEMs with $\SI{90}{\micro m}$ hole pitch and thus higher hole density have been used.
This allows a finer sampling of the primary ionisation electrons in the drift region and thus improves the preservation of the position information during the charge collection process.
As a result, the spatial resolution could be improved by up to $\sim\SI{15}{\percent}$.
Spatial resolutions down to $\SI{50}{\micro m}$ have been measured for the finer pitch geometry, by optimising the settings of the electric drift field and thus charge collection, using the centroid/COG method for position reconstruction.

Furthermore, the studies highlighted the impact of the front-end electronics on the spatial resolution determination.
Mainly in terms of saturated front-end channels decreasing the spatial resolution.
On the other hand, also the improvement of the spatial resolution when optimising the use of the dynamic range of the electronics.

In addition to the GEMs with $\SI{90}{\micro m}$ hole pitch, also GEMs with even smaller hole pitches of $\SI{60}{\micro m}$ and inner hole diameters of $\SI{25}{\micro m}$ exist.
These GEMs were tested in the scope of the here presented measurements, but require more thorough studies in the future.
Due to manufacturing requirements, the thickness of the GEM is reduced to $\SI{25}{\micro m}$, resulting in a different working point of the GEM, leading to discharges and detector instabilities.
Besides the even finer pitch GEMs, future studies may also focus on combining GEMs with higher hole density with readout geometries with finer strip pitch.

\section*{Acknowledgements}

This work has been sponsored by the Wolfgang Gentner Programme of the German Federal Ministry of Education and Research (grant no.\ 13E18CHA).

This work has been supported by the CERN EP R\&D Strategic Programme on Technologies for Future Experiments (\url{https://ep-rnd.web.cern.ch/}).

This project has received funding from the European Union's Horizon 2020 Research and Innovation programme under Grant Agreement No 101004761.

The authors would like to thank Jona Bortfeldt (LMU Munich) for providing the anamicom software (\url{https://gitlab.physik.uni-muenchen.de/Jonathan.Bortfeldt/anamicom}) to reconstruct the particle trajectories.

\end{document}